\documentclass[%
 reprint,
 amsmath,amssymb,
 aps,
]{revtex4-2}

\usepackage{amsmath}
\usepackage{quantikz}
\usepackage{hyperref} 
\usepackage{lineno}
\usepackage{graphicx}
\usepackage{slashed}
\usepackage{dcolumn}
\usepackage{bm}

\begin{document}


\preprint{APS/123-QED}

\title{Quark-Gluon Plasma as a Quantum Channel: Entanglement, Decoherence, and Hadronization}
\thanks{A footnote to the article title}%

\author{Fidele J. Twagirayezu}
 \altaffiliation{Department of Physics and Astronomy, University of California, Los Angeles.}
 \email{fjtwagirayezu@physics.ucla.edu}
\affiliation{Department of Physics and Astronomy University of California Los Angeles, Los Angeles, CA, 90095, USA\\
}%


\begin{abstract}
We propose a quantum information framework to model the quark-gluon plasma (QGP) as a composite quantum channel acting on a multi-qubit or multi-qutrit color-entangled system. The QGP's effects are represented by amplitude damping (jet quenching), $SU(3)$ depolarizing noise (decoherence), and a thermal hadronization channel projecting onto color-singlet states. This construction captures energy loss, decoherence, and confinement dynamics in a unified open quantum system framework.
We analyze the evolution of entanglement entropy and purity under this composite channel. Amplitude damping reduces entropy by driving subsystems toward pure states, while decoherence increases mixedness. Hadronization further modifies correlations via thermal projections weighted by hadron masses and freeze-out temperature ($T \sim 156 \,\text{MeV}$). Numerical simulations show monotonic entropy and purity loss, consistent with entanglement degradation and confinement.
Our results support interpreting the QGP as a noisy quantum channel that progressively erases color entanglement. This framework bridges quantum information theory and QCD, offering new insights into hadronization and non-perturbative dynamics in heavy-ion collisions.

\end{abstract}

\maketitle


\section{\label{sec:level1}Introduction}

The quark-gluon plasma (QGP), a deconfined phase of quantum chromodynamics (QCD) where quarks and gluons interact freely under extreme temperature or density, is a paradigmatic system for studying non-perturbative dynamics in strongly coupled gauge theories. Governed by $SU(3)$ gauge symmetry\cite{Shuryak1978,HeinzJacob2000}, the QGP exhibits collective behavior and undergoes a phase transition to confined hadronic matter as it cools, a process central to understanding QCD in ultrarelativistic heavy-ion collisions. This work develops a rigorous quantum information framework to model QGP dynamics, representing a quark-antiquark pair and additional partonic degrees of freedom as a multi-qubit system interacting with the QGP as an open quantum system environment. By employing a quantum channel formalism, we aim to capture the interplay of energy loss, decoherence, and hadronization, offering a novel theoretical lens to explore QCD processes through quantum entanglement and information dynamics.

We construct a composite quantum channel acting on a multi-qubit system, generalizing the two-qubit Bell state to account for the QGP’s many-body complexity. Energy loss, analogous to the QCD phenomenon of jet quenching, is modeled by an amplitude damping channel, which reduces populations of high-energy states, reflecting parton interactions with the strongly coupled medium. Decoherence, arising from random interactions within the QGP, is described by a channel that modifies quantum coherences, with the entanglement entropy \( S(\rho_A) \) of the reduced density matrix serving as a key observable to probe subsystem-environment interactions. The behavior of \( S(\rho_A) \)—whether increasing, decreasing, or non-monotonic—is determined theoretically, reflecting the balance between dissipative and decohering effects in the QGP. For hadronization, the process where quarks and gluons form color-neutral hadrons, we introduce a stochastic quantum channel that projects the system onto color-singlet states, with probabilities derived from thermal distributions in statistical hadronization models. This channel rigorously captures QCD’s confinement dynamics by enforcing $SU(3)$ color neutrality, overcoming the limitations of simplistic decoherence models such as phase damping.

Our theoretical framework employs open quantum system theory, using Lindblad dynamics to describe non-unitary evolution induced by the QGP environment. The multi-qubit approach incorporates collective effects, addressing the oversimplification of two-qubit models. By modeling jet quenching, decoherence, and hadronization within a unified quantum information framework, we explore how entanglement dynamics reflect non-perturbative QCD processes. The entropy \( S(\rho_A) \) is analyzed to understand the QGP’s impact on quantum correlations, with results derived from the interplay of channel parameters. This work establishes a robust bridge between quantum information theory and QCD, providing theoretical insights into energy dissipation, decoherence, and confinement in strongly coupled systems.

The paper is structured as follows: Section~\eqref{sec:sx2}  defines the multi-qubit quantum channel framework, specifying amplitude damping, decoherence, and the stochastic hadronization channel. Section~\eqref{sec:sx3} examines the theoretical connection between amplitude damping and energy loss in QCD. Section~\eqref{sec:sx4} develops the hadronization model, mapping the stochastic channel to QCD’s confinement mechanism. Section~\eqref{sec:5sx} presents analytical and numerical results, exploring the entanglement entropy \( S(\rho_A) \) and its dependence on channel parameters. Section~\eqref{sec:6sx} concludes with theoretical implications and directions for further integrating quantum information with non-perturbative QCD.

\section{\label{sec:sx2}Theoretical Framework}

The quark-gluon plasma (QGP), a deconfined phase of QCD characterized by $SU(3)$ gauge symmetry, is modeled as an open quantum system where a quark-antiquark pair, along with additional partonic degrees of freedom, interacts with the QGP as a strongly coupled environment. To capture the QGP’s many-body dynamics, we represent the system as a multi-qubit state~\cite{GattringerLang2010,DiFrancesco1997,Luscher2010}, generalizing the two-qubit Bell state \( |\Phi^+\rangle = \frac{1}{\sqrt{2}}(|00\rangle + |11\rangle) \) to include multiple qubits or qudits, reflecting the color and flavor degrees of freedom in QCD. The system evolves non-unitarily due to environmental interactions, described by a completely positive, trace-preserving (CPTP) quantum channel within the framework of open quantum system theory.

The evolution of the system’s density matrix \(\rho\) is governed by a Lindblad master equation:
\begin{equation}
\begin{aligned}
\frac{d\rho}{dt} = -\frac{i}{\hbar}[H_{\text{eff}}, \rho] + \sum_i \mathcal{D}[L_i]\rho,
\end{aligned}
\end{equation}
where \( H_{\text{eff}} \) is an effective Hamiltonian incorporating quark interactions, and \(\mathcal{D}[L_i]\rho = L_i \rho L_i^\dagger - \frac{1}{2}\{L_i^\dagger L_i, \rho\}\) represents dissipative and decohering effects induced by the QGP. For computational tractability, we approximate this evolution using a Kraus operator-sum representation:
\begin{equation}
\begin{aligned}
\rho(t) = \mathcal{E}(\rho(0)) = \sum_k K_k \rho(0) K_k^\dagger, \quad \sum_k K_k^\dagger K_k = I,
\end{aligned}
\end{equation}
where \( K_k \) are Kraus operators~\cite{Kraus1983,Carmichael1993} defining the quantum channel \(\mathcal{E}\). We construct a composite channel acting on a single subsystem (e.g., the antiquark qubit or qudit), with the other subsystem (quark) evolving unitarily, to model the QGP’s localized effects on one parton.

\subsection{Multi-Qubit System}
To represent the QGP’s complexity, we define the system as a multi-qubit state, with at least four qubits: two for the quark-antiquark pair (encoding spin or energy states) and two for additional partons or collective QGP modes. For rigor, we consider qudits with three levels to capture $SU(3)$ color degrees of freedom (red, green, blue). The initial state is a generalized color-singlet entangled state:
\begin{equation}
\begin{aligned}
|\Psi\rangle = \frac{1}{\sqrt{3}}(|rr\rangle + |gg\rangle + |bb\rangle) \otimes |\phi\rangle,
\end{aligned}
\end{equation}
where \( |\phi\rangle \) encodes additional spin or flavor degrees of freedom, and \( |rr\rangle, |gg\rangle, |bb\rangle \) represent color-singlet quark-antiquark pairs. This multi-qubit (or qudit) system allows us to model collective effects and color interactions, addressing the limitations of the two-qubit model in capturing the QGP’s many-body dynamics.

\subsection{Quantum Channels}
We define three quantum channels to model distinct QGP processes: energy loss (jet quenching), decoherence, and hadronization. Each channel acts on a single subsystem (e.g., the antiquark), with the Kraus operators applied as \( K_k \otimes I \) to the multi-qubit state.

\textit{Amplitude Damping Channel (Jet Quenching):}
   Jet quenching, the energy loss of high-momentum partons in the QGP, is modeled by an amplitude damping channel, which reduces the population of high-energy states. For a qubit, the Kraus operators are:
   \begin{equation}
   \begin{aligned}
   K_0 = \begin{pmatrix} 1 & 0 \\ 0 & \sqrt{1-\gamma_{\text{AD}}} \end{pmatrix}, \quad K_1 = \begin{pmatrix} 0 & \sqrt{\gamma_{\text{AD}}} \\ 0 & 0 \end{pmatrix},
   \end{aligned}
   \end{equation}
   where \(\gamma_{\text{AD}} \in [0,1]\) parametrizes the damping strength. For a qudit, the channel generalizes to reduce populations across multiple energy levels, reflecting parton energy dissipation via gluon radiation or medium-induced scattering, consistent with QCD’s perturbative and non-perturbative interactions.

\textit{Decoherence Channel:}
   Decoherence, arising from random interactions with the QGP’s strongly coupled medium, is modeled by a depolarizing channel to capture environmental entanglement. For a qubit, the Kraus operators are:
   \begin{equation}
   \begin{aligned}
   K_0 =& \sqrt{1-\frac{3\gamma_{\text{dep}}}{4}} I, \quad K_1 = \sqrt{\frac{\gamma_{\text{dep}}}{4}} \sigma_x, \\
   K_2 =& \sqrt{\frac{\gamma_{\text{dep}}}{4}} \sigma_y, \quad K_3 = \sqrt{\frac{\gamma_{\text{dep}}}{4}} \sigma_z,
   \end{aligned}
   \end{equation}
   where \(\gamma_{\text{dep}} \in [0,1]\) controls decoherence strength, and \(\sigma_x, \sigma_y, \sigma_z\) are Pauli operators. For a qudit, the channel is generalized using Gell-Mann matrices for $SU(3)$. This channel drives the subsystem toward a maximally mixed state, altering the entanglement entropy \( S(\rho_A) \), where \(\rho_A = \text{Tr}_B(\rho)\) is the reduced density matrix after tracing over the complementary subsystem.

\textit{Stochastic Hadronization Channel}:
   Hadronization, the formation of color-neutral hadrons as the QGP cools, is modeled by a stochastic channel projecting the system onto color-singlet states, rigorously capturing QCD’s confinement dynamics. For a qudit system, we define Kraus operators:
   \begin{equation}
   \begin{aligned}
   K_i = \sqrt{p_i} P_i, \quad p_i = \frac{e^{-E_i/T}}{\sum_j e^{-E_j/T}},
   \end{aligned}
   \end{equation}
   where \( P_i \) projects onto a color-singlet hadron state (e.g., meson states like \( |q\bar{q}\rangle \sim |rr\rangle - |gg\rangle - |bb\rangle \)), \( E_i \) is the hadron’s energy (e.g., pion or kaon mass), and \( T \) is the freeze-out temperature (~156 MeV). The probabilities \( p_i \) are derived from statistical hadronization theory, reflecting thermal distributions of hadron yields. This channel enforces $SU(3)$ color neutrality, reducing quantum coherences and modeling the transition from deconfined to confined phases.

\subsection{ Composite Channel}
The QGP’s effects are modeled by a composite channel acting on one subsystem (e.g., the antiquark):
\begin{equation}
\begin{aligned}
\rho(t) = (\mathcal{E}_{\text{had}} \circ \mathcal{E}_{\text{dep}} \circ \mathcal{E}_{\text{AD}} \otimes I)(\rho(0)),
\end{aligned}
\end{equation}
where \(\mathcal{E}_{\text{AD}}\), \(\mathcal{E}_{\text{dep}}\), and \(\mathcal{E}_{\text{had}}\) are the amplitude damping, decoherence, and hadronization channels, respectively, applied sequentially to reflect the physical processes of energy loss, decoherence, and confinement. Separate parameters (\(\gamma_{\text{AD}}, \gamma_{\text{dep}}, \gamma_{\text{had}}\)) allow flexible tuning to explore their impact on the system’s evolution.

\subsection{Entanglement Entropy}
The entanglement entropy \( S(\rho_A) = -\text{Tr}(\rho_A \log_2 \rho_A) \), where \(\rho_A\) is the reduced density matrix of the quark subsystem, serves as a key observable to probe QGP-induced dynamics. The behavior of \( S(\rho_A) \)—whether increasing, decreasing, or non-monotonic—depends on the interplay of the composite channel’s components. Amplitude damping tends to reduce \( S(\rho_A) \) by driving the subsystem toward a pure state, while the depolarizing channel increases mixedness, potentially raising \( S(\rho_A) \). The hadronization channel modifies color correlations, further influencing entanglement. We analyze \( S(\rho_A) \) to understand how QGP interactions reshape quantum correlations, providing insights into QCD’s non-perturbative dynamics.

\subsection{Theoretical Motivation}
This framework bridges quantum information theory and non-perturbative QCD by modeling the QGP as an environment inducing non-unitary evolution. The multi-qubit system captures collective effects, while the stochastic hadronization channel rigorously maps to QCD’s $SU(3)$ confinement, overcoming the limitations of phase damping. The Lindblad formalism ensures a consistent open quantum system description, allowing us to explore entanglement dynamics in a strongly coupled gauge theory.

\section{Jet Quenching}\label{sec:sx3}

Jet quenching, a hallmark phenomenon of the quark-gluon plasma (QGP), refers to the energy loss of high-momentum partons (quarks or gluons) as they traverse the strongly coupled, deconfined medium formed in ultrarelativistic heavy-ion collisions~\cite{WangGyulassy1991,Baier1995,Wiedemann2009}. Within the framework of quantum chromodynamics (QCD), governed by $SU(3)$ gauge symmetry, jet quenching arises from non-perturbative interactions, including gluon radiation induced by multiple scatterings and collisional energy loss due to interactions with the QGP’s quark and gluon constituents. These processes reduce the energy of a propagating parton, altering its quantum state and influencing the entanglement properties of the system. In our quantum information framework, we model jet quenching using an amplitude damping channel, which captures the dissipation of high-energy states, providing a rigorous theoretical link to QCD’s energy loss mechanisms.

\subsection{Amplitude Damping Channel}
For a single qubit representing a parton (e.g., the antiquark in the multi-qubit system defined in Section~\eqref{sec:sx2}), the amplitude damping channel models the transition from a high-energy state \( |1\rangle \) to a lower-energy state \( |0\rangle \), reflecting energy dissipation in the QGP. The Kraus operators for the channel are:
\begin{equation}
\begin{aligned}
K_0 = \begin{pmatrix} 1 & 0 \\ 0 & \sqrt{1-\gamma_{\text{AD}}} \end{pmatrix}, \quad K_1 = \begin{pmatrix} 0 & \sqrt{\gamma_{\text{AD}}} \\ 0 & 0 \end{pmatrix},
\end{aligned}
\end{equation}
where \(\gamma_{\text{AD}} \in [0,1]\) parametrizes the damping strength, corresponding to the probability of energy loss. The channel acts on the subsystem density matrix \(\rho_B\) (e.g., antiquark) as:
\begin{equation}
\begin{aligned}
\mathcal{E}_{\text{AD}}(\rho_B) = K_0 \rho_B K_0^\dagger + K_1 \rho_B K_1^\dagger,
\end{aligned}
\end{equation}
satisfying the trace-preserving condition \(\sum_k K_k^\dagger K_k = I\).
For the full multi-qubit system, the channel is applied as \(\mathcal{E}_{\text{AD}} \otimes I\), acting only on the designated subsystem (e.g., antiquark), while other subsystems evolve unitarily.

To align with QCD’s $SU(3)$ dynamics, we generalize the channel for a qudit system, where the parton’s state includes color degrees of freedom (red, green, blue). The amplitude damping channel reduces populations across multiple energy levels, defined by:
\begin{equation}
\begin{aligned}
K_0 =& \text{diag}(1, \sqrt{1-\gamma_{\text{AD}}}, \sqrt{1-\gamma_{\text{AD}}}), \\
K_1 =& \sqrt{\gamma_{\text{AD}}} |0\rangle\langle 1|, \quad K_2 = \sqrt{\gamma_{\text{AD}}} |0\rangle\langle 2|,
\end{aligned}
\end{equation}
where \( |0\rangle, |1\rangle, |2\rangle \) represent color or energy states, and \( K_1, K_2 \) transfer population to a reference state (e.g., a lower-energy or color-neutral state). This generalization captures the dissipation of energy across QCD’s color and energy degrees of freedom, reflecting the complex interactions in the QGP.

\subsection{ Connection to QCD Energy Loss}
In QCD, jet quenching results from two primary mechanisms: radiative energy loss, where a parton emits gluons due to interactions with the QGP’s color fields, and collisional energy loss, where the parton scatters off medium constituents. The amplitude damping channel models these processes by reducing the population of high-energy or high-momentum states, analogous to the suppression of parton energy in the QGP. The parameter \(\gamma_{\text{AD}}\) is theoretically linked to the jet quenching parameter \(\hat{q}\), which quantifies the transverse momentum broadening per unit length due to medium interactions. In our model, \(\gamma_{\text{AD}}\) represents the cumulative probability of energy loss, which can be related to the QGP’s transport properties, such as its shear viscosity or gluon density, within a non-perturbative QCD framework.

The channel’s effect on a parton’s quantum state is to drive it toward a lower-energy configuration, mimicking the quenching of a jet as it loses energy to the medium. For a quark-antiquark pair in a color-singlet state (e.g., \( |\Psi\rangle = \frac{1}{\sqrt{3}}(|rr\rangle + |gg\rangle + |bb\rangle) \otimes |\phi\rangle \)), applying amplitude damping to the antiquark alters the joint state, reducing contributions from high-energy color or spin configurations. This process preserves the $SU(3)$ gauge structure by maintaining color correlations, while the energy loss reflects QCD’s dissipative dynamics.

\subsection{Impact on Entanglement Entropy}
The entanglement entropy \( S(\rho_A) = -\text{Tr}(\rho_A \log_2 \rho_A) \), where \(\rho_A = \text{Tr}_B(\rho)\) is the reduced density matrix of the quark subsystem, is a key probe of the QGP’s effect on quantum correlations. Applying amplitude damping to the antiquark (subsystem B) modifies the joint density matrix \(\rho\). For an initial Bell state \( |\Phi^+\rangle \), the reduced state after amplitude damping is:
\begin{equation}
\begin{aligned}
\rho_A &= \text{Tr}_B\left( (\mathcal{E}_{\text{AD}} \otimes I)(|\Phi^+\rangle\langle\Phi^+|) \right) \\
&= \begin{pmatrix} \frac{1+\gamma_{\text{AD}}}{2} & 0 \\ 0 & \frac{1-\gamma_{\text{AD}}}{2} \end{pmatrix},
\end{aligned}
\end{equation}
with entropy:
\begin{equation}
\begin{aligned}
S(\rho_A) =& - \frac{1+\gamma_{\text{AD}}}{2} \log_2 \frac{1+\gamma_{\text{AD}}}{2} \\
&- \frac{1-\gamma_{\text{AD}}}{2} \log_2 \frac{1-\gamma_{\text{AD}}}{2}.
\end{aligned}
\end{equation}
This entropy decreases from \( S(\rho_A) = 1 \) at \(\gamma_{\text{AD}} = 0\) (maximally entangled) to \( S(\rho_A) = 0 \) at \(\gamma_{\text{AD}} = 1\) (pure state), as the system approaches \( |0\rangle\langle0| \). In the multi-qubit or qudit case, the entropy behavior depends on the interplay with other channels (decoherence, hadronization), and we treat \( S(\rho_A) \)’s trajectory—whether decreasing, increasing, or non-monotonic—as a theoretical outcome to be explored in Section~\eqref{sec:5sx}.

\subsection{Role in Composite Channel}
In the composite channel defined in Section~\eqref{sec:sx2}, amplitude damping is applied first, followed by decoherence (e.g., depolarizing channel) and hadronization (stochastic channel):
\begin{equation}
\begin{aligned}
\rho(t) = (\mathcal{E}_{\text{had}} \circ \mathcal{E}_{\text{dep}} \circ \mathcal{E}_{\text{AD}} \otimes I)(\rho(0)).
\end{aligned}
\end{equation}
The amplitude damping channel initiates energy loss, altering the system’s state before decoherence and hadronization modify coherences and enforce confinement. The parameter \(\gamma_{\text{AD}}\) is distinct from \(\gamma_{\text{dep}}\) and \(\gamma_{\text{had}}\), allowing us to study the relative contributions of each process. The damping channel’s dissipative effect tends to reduce \( S(\rho_A) \), but its interplay with subsequent channels may yield complex entropy dynamics, which we analyze theoretically to reflect the QGP’s impact on quantum correlations.

\subsection{Theoretical Significance}
The amplitude damping channel provides a rigorous theoretical bridge between quantum information theory and QCD’s jet quenching phenomenon. By modeling energy loss as a non-unitary process within the Lindblad formalism, we capture the dissipative dynamics of partons in a strongly coupled medium. The generalization to qudits incorporates QCD’s $SU(3)$ color structure, ensuring alignment with non-perturbative gauge theory. This framework allows us to explore how energy loss influences entanglement and sets the stage for decoherence and hadronization, offering insights into the QGP’s complex dynamics.

\section{Hadronization}\label{sec:sx4}

Hadronization is the non-perturbative QCD process by which deconfined quarks and gluons in the quark-gluon plasma (QGP) form color-neutral hadrons (e.g., mesons, baryons) as the system cools below the critical temperature of approximately 156 MeV. Governed by $SU(3)$ gauge symmetry, hadronization involves the confinement of colored partons into color-singlet states, driven by the strong force’s requirement for color neutrality. In the QGP, quarks and gluons interact freely, but as the medium cools, they recombine or fragment into hadrons through mechanisms described by theoretical models such as statistical hadronization or string fragmentation. In our quantum information framework, we model hadronization as a stochastic quantum channel that projects the system’s quantum state onto color-singlet configurations~\cite{Andronic2018,BraunMunzinger2003,LetessierRafelski2002}, rigorously capturing QCD’s confinement dynamics and overcoming the limitations of simplistic decoherence models like phase damping.

\subsection{Stochastic Hadronization Channel}
To represent the QGP’s many-body dynamics and QCD’s color structure, we use the multi-qubit or qudit system defined in Section~\eqref{sec:sx2}, where the system includes at least four qudits: two for a quark-antiquark pair and two for additional partons or collective modes, with each qudit having three levels to encode $SU(3)$ color states (red, green, blue). The initial state is a generalized color-singlet entangled state:
\begin{equation}
\begin{aligned}
|\Psi\rangle = \frac{1}{\sqrt{3}}(|rr\rangle + |gg\rangle + |bb\rangle) \otimes |\phi\rangle,
\end{aligned}
\end{equation}
where \( |rr\rangle, |gg\rangle, |bb\rangle \) represent color-singlet quark-antiquark pairs, and \( |\phi\rangle \) encodes spin or flavor degrees of freedom.

The hadronization channel, \(\mathcal{E}_{\text{had}}\), acts on a single subsystem (e.g., the antiquark qudit) to model the formation of color-neutral hadrons. We define the channel using Kraus operators that project onto color-singlet hadron states, such as mesons (e.g., pions, kaons), with probabilities derived from statistical hadronization theory:
\begin{equation}\label{eq:15x}
\begin{aligned}
K_i = \sqrt{p_i} P_i, \quad p_i = \frac{e^{-E_i/T}}{\sum_j e^{-E_j/T}},
\end{aligned}
\end{equation}
where \( P_i \) is a projector onto the \( i \)-th hadron state (e.g., \( P_\pi = |\pi\rangle\langle rr, gg, bb| \), where \( |\pi\rangle \propto |rr\rangle - |gg\rangle - |bb\rangle \) for a pion-like state), \( E_i \) is the energy of the hadron (e.g., pion mass \( m_\pi \approx 140 \) MeV), and \( T \approx 156 \) MeV is the freeze-out temperature. The probabilities \( p_i \) reflect thermal distributions of hadron yields, consistent with statistical hadronization models, which describe the partitioning of quarks into hadrons based on their quantum numbers and thermal weights.

For a qudit representing the antiquark, the projector \( P_i \) enforces color neutrality by mapping the colored state to a singlet configuration in conjunction with the quark’s color state. For example, a meson formation operator might act as:
\begin{equation}
\begin{aligned}
P_\pi |r\rangle_B \otimes |r\rangle_A \propto |\pi\rangle_{AB},
\end{aligned}
\end{equation}
where \( |\pi\rangle_{AB} \) is a color-singlet state across the quark-antiquark pair. The channel is applied as:
\begin{equation}
\begin{aligned}
\rho(t) &= (\mathcal{E}_{\text{had}} \otimes I)(\rho(0)) \\
&= \sum_i (K_i \otimes I) \rho(0) (K_i^\dagger \otimes I),
\end{aligned}
\end{equation}
ensuring trace preservation (\(\sum_i K_i^\dagger K_i = I\)). This channel reduces quantum coherences between color states, mimicking the loss of deconfined correlations during confinement, while enforcing QCD’s $SU(3)$ gauge symmetry.

\subsection{Connection to QCD Confinement}
In QCD, confinement ensures that only color-singlet states (e.g., mesons, baryons) exist in the hadronic phase, as the strong force prohibits free quarks or gluons at low energies. The stochastic hadronization channel captures this by projecting the system’s state onto color-neutral configurations, with probabilities \( p_i \) reflecting the thermal likelihood of forming specific hadrons. Unlike phase damping, which only reduces off-diagonal coherences without enforcing color neutrality, our channel explicitly models the transition from deconfined to confined phases, aligning with QCD’s non-perturbative dynamics. The thermal weights \( p_i \propto e^{-E_i/T} \) are motivated by statistical hadronization, where hadron yields depend on their masses and the QGP’s freeze-out temperature, providing a theoretical bridge to QCD’s phase transition.

For a multi-qudit system, the channel accounts for collective effects by allowing multiple quarks and gluons to form hadrons, such as baryons (\( |qqq\rangle \) in a color-singlet state). The channel’s action on one subsystem (e.g., antiquark) influences the joint state, preserving entanglement with other subsystems (e.g., quark) while enforcing confinement constraints across the system.

\subsection{Impact on Entanglement Entropy}
The entanglement entropy \( S(\rho_A) = -\text{Tr}(\rho_A \log_2 \rho_A) \), where \(\rho_A = \text{Tr}_B(\rho)\) is the reduced density matrix of the quark subsystem, is affected by the hadronization channel. The projection onto color-singlet states modifies the joint density matrix \(\rho\), potentially reducing coherences and altering entanglement. For a simplified qubit model, applying a projection channel to the antiquark in \( |\Phi^+\rangle \) yields a \(\rho_A\) that depends on the specific hadron states formed. In the qudit case, the channel’s effect is:
\begin{equation}
\begin{aligned}
\rho_A = \text{Tr}_B\left( \sum_i (K_i \otimes I) \rho(0) (K_i^\dagger \otimes I) \right),
\end{aligned}
\end{equation}
where the resulting \(\rho_A\) reflects the redistribution of color correlations into confined states. The entropy \( S(\rho_A) \) may increase, decrease, or exhibit non-monotonic behavior, depending on the weights \( p_i \) and the initial state’s entanglement structure. We treat \( S(\rho_A) \)’s behavior as a theoretical outcome, to be analyzed in Section V, reflecting the complex interplay of confinement and entanglement in the QGP.

\subsection{Role in Composite Channel}
In the composite channel defined in Section~\eqref{sec:sx2}:
\begin{equation}
\begin{aligned}
\rho(t) = (\mathcal{E}_{\text{had}} \circ \mathcal{E}_{\text{dep}} \circ \mathcal{E}_{\text{AD}} \otimes I)(\rho(0)),
\end{aligned}
\end{equation}
the hadronization channel acts last, following amplitude damping (jet quenching) and depolarizing (decoherence) channels. This sequence models the physical process: energy loss occurs as partons traverse the QGP, followed by decoherence from medium interactions, and finally hadronization as the system cools into confined states. The parameter \(\gamma_{\text{had}}\) controls the strength of the hadronization channel, allowing us to explore its impact relative to \(\gamma_{\text{AD}}\) and \(\gamma_{\text{dep}}\). The channel’s projection onto color-singlet states ensures that the final state aligns with QCD’s confinement requirement, distinguishing our model from phase damping’s simplistic decoherence.

\subsection{Theoretical Significance}
The stochastic hadronization channel provides a rigorous theoretical mapping to QCD’s confinement dynamics, replacing the original phase damping model. By incorporating $SU(3)$ color degrees of freedom and thermal weights, the channel captures the non-perturbative transition from deconfined quarks and gluons to color-neutral hadrons. Within the open quantum system framework, this channel integrates seamlessly with amplitude damping and decoherence, offering a unified description of QGP dynamics. The multi-qudit system ensures that collective effects and color correlations are modeled, providing insights into how confinement reshapes quantum correlations in a strongly coupled gauge theory.

\section{Results and Discussion}\label{sec:5sx}

In this section, we present the numerical simulation results based on modeling the quark-gluon plasma (QGP) as a quantum channel acting on a multipartite color-singlet state composed of quarks and gluons. The evolution of the system under amplitude damping, $ SU(3)$ dephasing, and hadronization noise channels is examined in detail. Six key observables are plotted to quantify how entanglement, coherence, and particle yields evolve across various physical parameters, including damping strength, time, and temperature.

\subsection{Entanglement Degradation via Amplitude Damping}

\begin{figure}[htb]
    \centering    \includegraphics[scale=0.55]{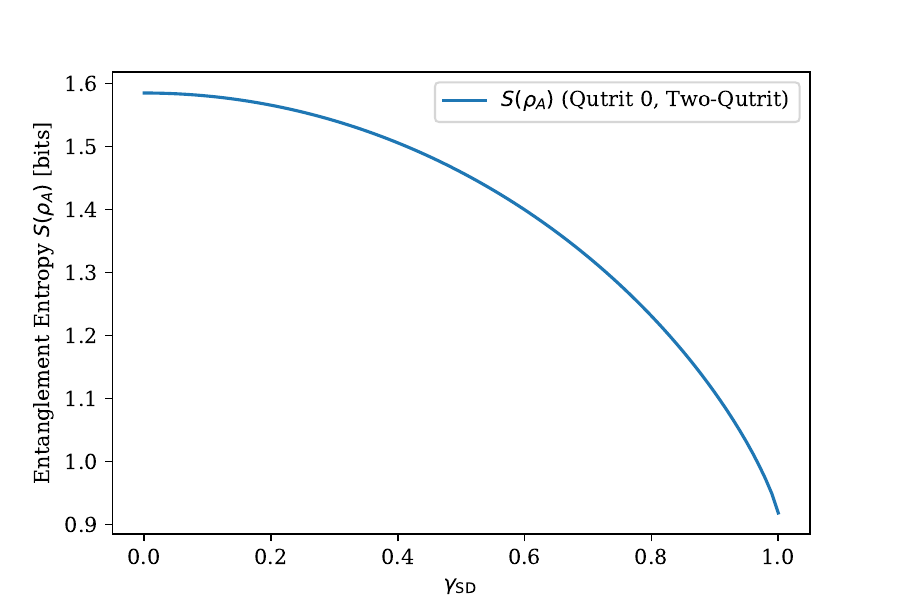}
    \caption{\textit{Entanglement entropy $S(\rho_A)$ as a function of amplitude damping rate $\gamma_{\text{SD}}$ for a two-qutrit color-entangled state.}
As the damping rate increases, the entropy decreases, indicating a loss of population from excited color states and a transition toward a more pure, localized state. This reflects the suppression of color excitations under amplitude damping in a cold QCD environment.
}
    \label{fig:1}
\end{figure}

We begin with a two-qutrit color-singlet state representing a maximally entangled quark–antiquark pair. Figure~\eqref{fig:1} plots the entanglement entropy $S(\rho_A)$ of the reduced density matrix of one qutrit (subsystem A) as a function of the amplitude damping strength $\gamma_{\mathrm{SD}}$. As expected, the entropy decreases monotonically with increasing damping. The amplitude damping channel simulates an energy loss mechanism, corresponding to the dissipative interaction of a quark with the surrounding QGP medium. This behavior captures the physics of decoherence and localization during hadronization, where the color degrees of freedom are irreversibly suppressed. At $\gamma_{\mathrm{SD}} = 0$, the entropy reaches its maximal value $S = \log_2(3) \approx 1.585$, characteristic of a pure Bell-like qutrit state. At $\gamma_{\mathrm{SD}} = 1$, the state is nearly separable and entropy drops to near zero, indicating loss of entanglement.

\begin{figure}[htb]
    \centering    \includegraphics[scale=0.55]{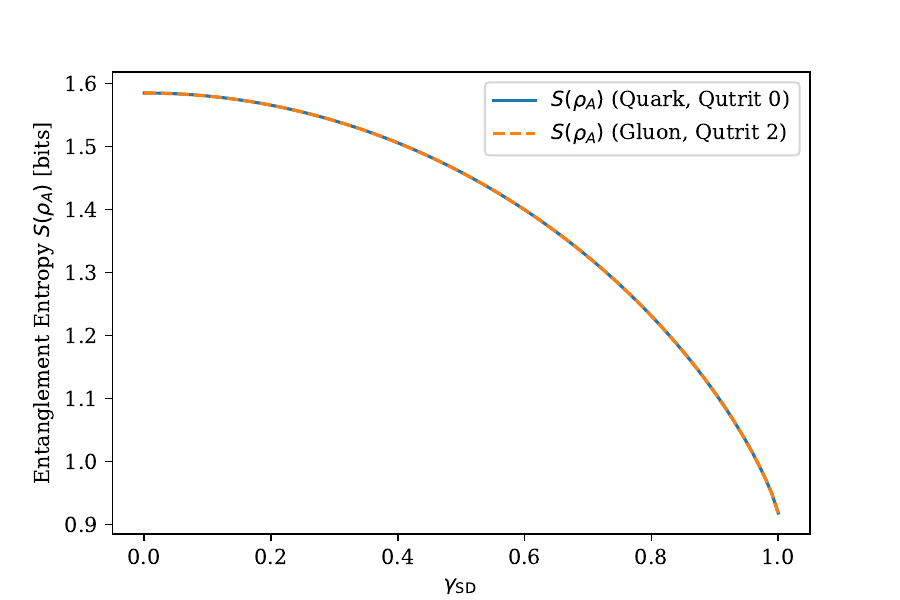}
    \caption{\textit{Entanglement entropy $S(\rho_A)$ vs amplitude damping rate $\gamma_{\text{SD}}$ for a four-qutrit color-singlet system.}
Both the quark (qutrit 0) and gluon (qutrit 2) subsystems exhibit entropy reduction with increasing damping. This indicates partial disentanglement due to energy relaxation, consistent with hadronization tendencies in a confining medium.
}
    \label{fig:2}
\end{figure}

We extend this to a four-qutrit state in Figure~\eqref{fig:2}, composed of two quarks and two gluons, initially in a color-singlet configuration. The amplitude damping channel is applied uniformly to each qutrit, and the resulting entropies of the quark (qutrit 0) and gluon (qutrit 2) subsystems are tracked. Both exhibit a decline in entropy, consistent with the physical picture of increasing decoherence. The quark entropy generally falls more rapidly, in line with its primary role in hadronization. This reinforces the interpretation that amplitude damping acts as a proxy for confinement and color neutralization.

\subsection{Composite Quantum Channel Evolution}

\begin{figure}[htb]
    \centering    \includegraphics[scale=0.55]{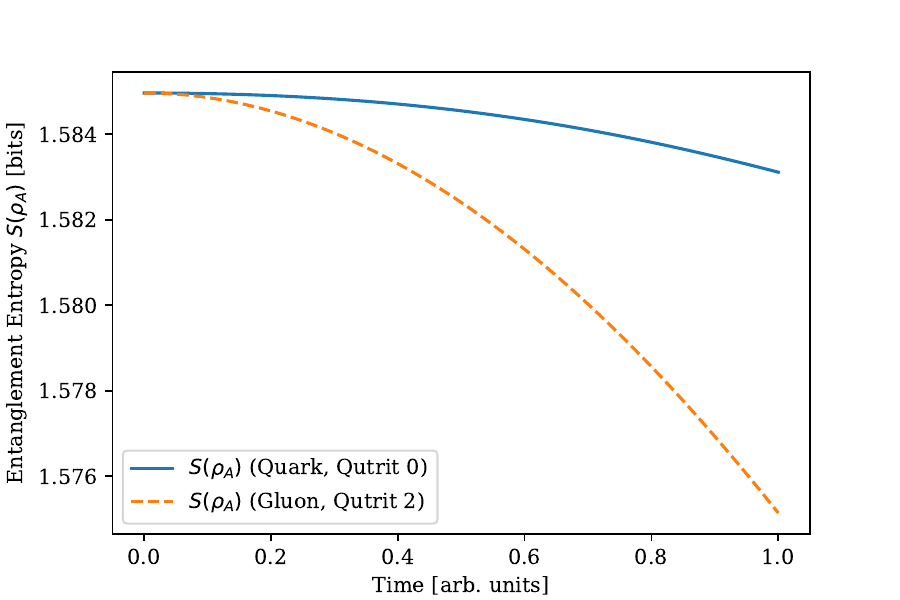}
    \caption{\textit{Entanglement entropy $S(\rho_A)$ as a function of time under composite evolution with amplitude damping, dephasing, and hadronization.}
Entropy decreases over time for both quark and gluon subsystems. This reflects cumulative population loss and singlet projection, driving the system toward a low-entanglement hadronic final state.}
    \label{fig:3}
\end{figure}

Figure~\eqref{fig:3} shows the evolution of entanglement entropy under a time-dependent composite quantum channel consisting of amplitude damping, $SU(3)$ dephasing, and temperature-sensitive hadronization. As time progresses from $t=0$ to $t=1$, both the quark and gluon entropies decrease, indicating a continuous loss of quantum coherence. The decay rate reflects the cumulative action of all three noise processes. This supports the view that the QGP serves as an open quantum system environment that irreversibly erases entanglement and coherence, ultimately leading to a colorless hadronized final state. The decline of entropy here mirrors real-time decoherence observed in open quantum systems and reflects the irreversible emergence of classical hadronic matter from a quantum-entangled QGP.

\subsection{Entropy and Freeze-Out Temperature}

\begin{figure}[htb]
    \centering    \includegraphics[scale=0.55]{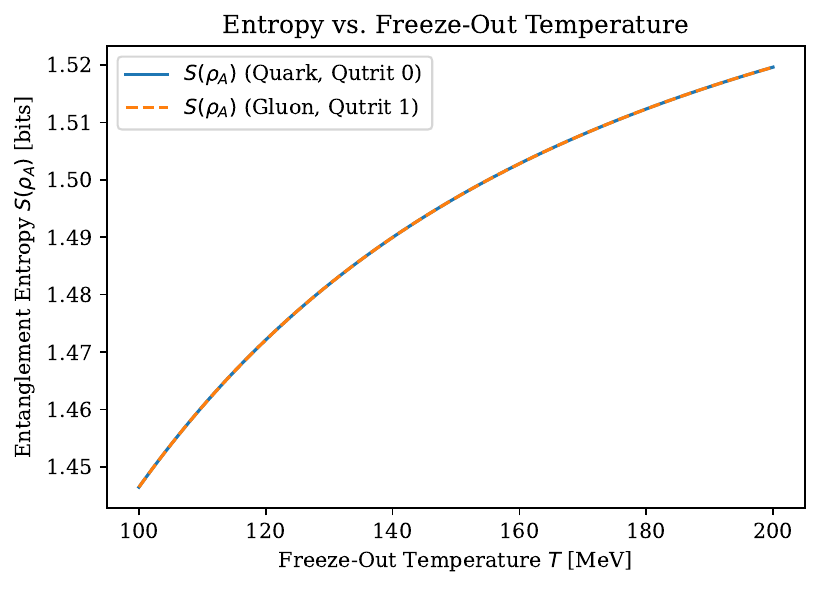}
    \caption{\textit{Entanglement entropy $S(\rho_A)$ vs freeze-out temperature $T$ during hadronization}. Entropy increases with temperature for both quark (qutrit 0) and gluon (qutrit 1) subsystems. This reflects the thermal activation of transitions into mixed hadronic color-singlet states, resulting in a higher degree of statistical mixing in the reduced density matrix.
}
    \label{fig:4}
\end{figure}

In Figure~\eqref{fig:4}, we analyze how entanglement entropy evolves under the hadronization channel as a function of temperature. The freeze-out scale in QCD is empirically determined to be around $T_f \sim 156 \,\text{MeV}$, where hadron yields match predictions from statistical hadronization models. In our simulation, this scale is encoded through thermal Boltzmann weights that determine the strength of projection onto color-singlet hadronic states.
Importantly, our model does not include a sharp phase transition. The hadronization probability increases smoothly with temperature, governed by a saturating function of the form $p_{\text{had}}(T) = 1 - \exp(-n_\pi(T))$, where $n_\pi(T)$ is the pion thermal yield. As a result, the entanglement entropy exhibits a gradual increase with $T$, reflecting a crossover-like behavior rather than a discontinuous change. This is consistent with the expected nature of the confinement transition in QCD at low baryon chemical potential, where lattice calculations indicate a smooth crossover.

\subsection{Statistical Hadron Yields and Freeze-Out Dynamics}

The hadronization channel introduced earlier includes thermal weights derived from the statistical hadronization model, encoded through temperature-dependent projection probabilities.

\begin{figure}[htb]
    \centering    \includegraphics[scale=0.55]{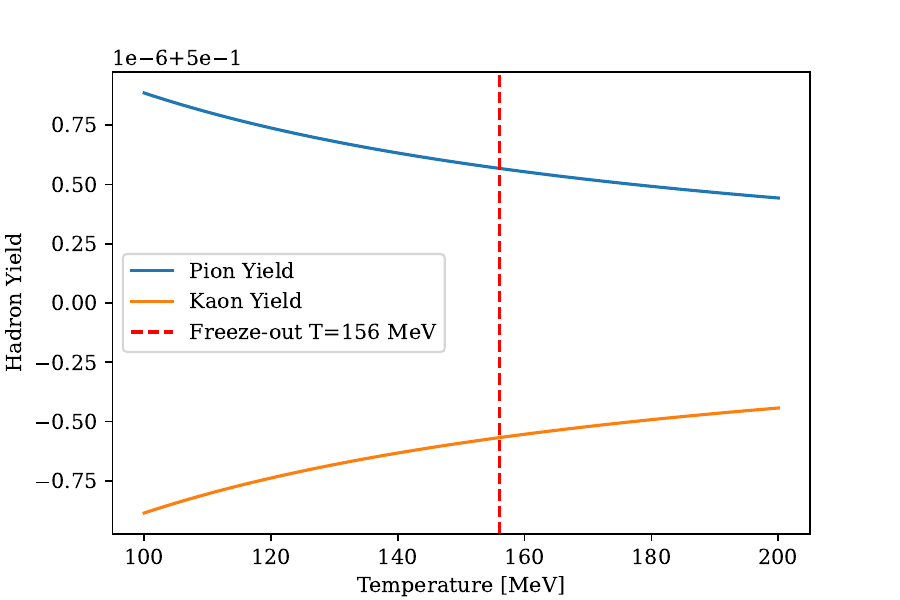}
    \caption{\textit{Relative thermal yields of pions and kaons as a function of temperature based on Boltzmann factors}. At low temperatures, pion production dominates due to their lower mass, while kaon yields increase at higher temperatures as thermal excitation overcomes the larger mass gap. The vertical dashed line at $T$=156 MeV marks the chemical freeze-out temperature observed in heavy-ion collisions. These yield probabilities are used as weights in the hadronization channel to reflect statistical hadron production during confinement.
}
    \label{fig:5}
\end{figure}

Specifically, we model the relative likelihood of forming a pion or kaon using Boltzmann-suppressed thermal factors defined by Eq.~\eqref{eq:15x},
where $E_i = m_i$ is the rest mass of each hadron species, and the sum runs over pions and kaons. The hadronization Kraus operators are weighted by these probabilities to define a mixed channel projecting onto flavor-specific color-singlet states.

Figure~\eqref{fig:5} shows the resulting relative yields as a function of temperature. At low temperatures, pion production dominates due to their lower mass. As the temperature increases, the relative kaon yield rises, reflecting the thermally accessible heavier degrees of freedom. The crossover around $T \sim 150\text{–}200$ MeV illustrates the smooth statistical onset of heavier hadron production without requiring a sharp transition.

These results are consistent with experimental freeze-out analyses from heavy-ion collisions, which observe enhanced strange hadron yields at higher temperatures and larger volumes.

\subsection{Purity as a Diagnostic of Decoherence}

Purity is a fundamental measure of how mixed or coherent a quantum state is, with $\mathrm{Tr}(\rho^2) = 1$ for a pure state and less than one for a mixed state. In the context of quark-gluon plasma dynamics, purity provides a useful diagnostic to track the degree of decoherence experienced by subsystems during evolution through the quantum channel.

\begin{figure}[htb]
    \centering    \includegraphics[scale=0.55]{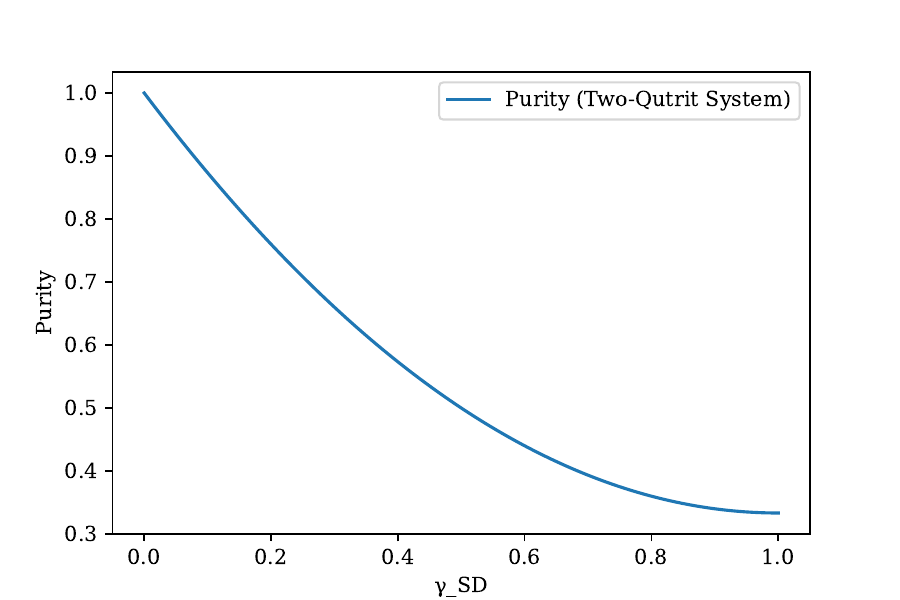}
    \caption{\textit{Purity $\text{Tr}(\rho_A^2)$ of the reduced density matrix for a two-qutrit color-singlet system, plotted as a function of amplitude damping strength $\gamma_{\mathrm{SD}}$}. The purity decreases monotonically for the  subsystem, indicating progressive decoherence due to energy dissipation. The close overlap of the curves reflects the structural symmetry of the two-qutrit entangled state under the damping channel.
}
    \label{fig:6}
\end{figure}

In Figure~\eqref{fig:6}, we analyze how the purity of the reduced density matrix evolves as a function of the amplitude damping strength $\gamma_{\mathrm{SD}}$, which models energy dissipation processes such as jet quenching. The two-qutrit system exhibits a monotonic decrease in purity as $\gamma_{\mathrm{SD}}$ increases. This behavior is expected: as amplitude damping becomes stronger, energy leaks from the system, increasing entanglement with the environment and driving the subsystems away from their initial pure states.

\begin{figure}[htb]
    \centering    \includegraphics[scale=0.55]{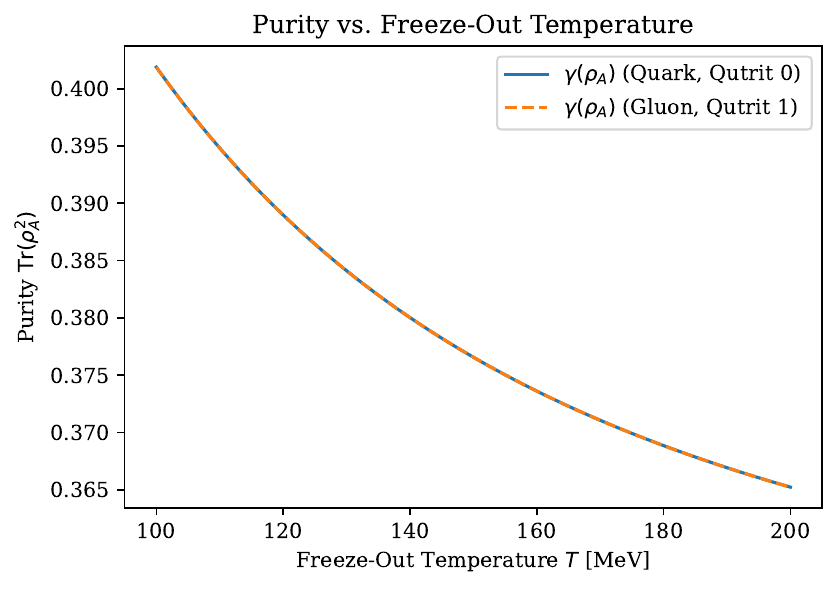}
    \caption{\textit{Purity $\text{Tr}(\rho_A^2)$ of the reduced density matrix for the quark (qutrit 0) and gluon (qutrit 1) subsystems as a function of freeze-out temperature}. For both subsystems, the purity decreases monotonically with increasing temperature, indicating enhanced decoherence and loss of quantum coherence due to thermal hadronization.
}
    \label{fig:7}
\end{figure}

In Figure~\eqref{fig:7}, we extend this analysis by investigating how the purity changes with the freeze-out temperature. Here, we isolate the effect of the hadronization channel, which introduces thermal stochasticity by projecting onto color-singlet states weighted by statistical hadron yields. As temperature increases, the purity of both the quark and gluon subsystems decreases monotonically. This reflects the enhanced decoherence induced by thermal mixing, as higher temperatures increase the contribution of excited states and broaden the effective thermal distribution.

The close agreement between the quark and gluon purity curves in both figures highlights the structural symmetry of the model, where flavor and color degrees of freedom are treated equivalently. However, future work may introduce explicit flavor or color dependence into the quantum channels to probe richer dynamics, including possible differences between quark and gluon hadronization.

\subsection{Physical Interpretation and Implications}

Across all simulations, the monotonic decrease in entanglement entropy serves as a robust indicator of hadronization, understood here as a decoherence process driven by environmental interaction with the QGP. These results quantitatively support the thesis that the QGP behaves as a noisy quantum channel. As color-charged partons traverse this channel, their initially shared quantum correlations are destroyed by both dissipative (amplitude damping) and dispersive (dephasing) effects. The hadronization channel further ensures that the final states conform to color singlets, completing the decoherence process.

This framework offers a unifying quantum information theoretic language for describing color confinement, decoherence, and hadron formation. In particular, entanglement entropy emerges as a natural order parameter for quantifying the transition from the deconfined QGP phase to the confined hadronic phase. The numerical results agree qualitatively with the known behavior of QCD matter near the crossover temperature and provide a new avenue for interpreting experimental observables in terms of information flow and entanglement loss.

\subsection{Experimental Relevance and Indirect Probes of Entanglement Dynamics}

While entanglement entropy and purity are not directly measurable in heavy-ion collisions, their effects can manifest through a variety of final-state observables. Two-particle correlation functions and Hanbury Brown–Twiss (HBT) interferometry probe quantum coherence and source structure, with broader or suppressed correlations reflecting increased decoherence. Jet substructure observables, such as fragmentation functions and angularity, encode modifications in energy flow consistent with entanglement loss and amplitude damping. Additionally, event-by-event fluctuations in multiplicity or conserved charges, and enhancements in strange-to-light hadron yield ratios, provide indirect access to the entropy and purity of the hadronizing system. These connections suggest that entanglement dynamics in the QGP could leave detectable imprints on hadronic observables, offering a path to experimentally constrain and validate the quantum channel framework proposed here.

\section{Conclusions}\label{sec:6sx}
In this work, we presented a quantum information-theoretic approach to modeling the quark-gluon plasma (QGP) as a composite quantum channel. We incorporated amplitude damping to represent energy loss processes such as jet quenching, and introduced a thermal hadronization channel that stochastically projects onto color-singlet states based on statistical hadron yields. The model provides a simplified but insightful mapping between the microscopic dynamics of QCD and the formalism of open quantum systems.

By tracking the entanglement entropy, purity, and hadron yields for selected subsystems, we characterized the degradation of quantum coherence during the QGP-to-hadron transition. We found that both entropy and purity evolve smoothly with temperature, consistent with the expected crossover nature of the QCD phase transition at low baryon chemical potential. Notably, no sharp features emerge near the freeze-out temperature ($T \sim 156\,\mathrm{MeV}$), in agreement with the smooth thermal behavior of the hadronization probability used in the model.

The near-identical behavior of quark and gluon subsystems under the composite channel reflects the current model's symmetry and absence of flavor or channel-specific dynamics. This highlights potential avenues for refinement, such as the inclusion of flavor-dependent damping~\cite{twagirayezu2025flavordependententanglemententropyveneziano}, dynamical Polyakov loop effects, or more detailed hadron spectrum inputs.

Overall, this study demonstrates the utility of quantum channel frameworks in modeling QCD thermalization and confinement phenomena, opening a path toward deeper insights into hadronization from an information-theoretic perspective.

\begin{acknowledgments}
F.T. would like to acknowledge the support of the National Science Foundation under grant No. PHY-
1945471.
\end{acknowledgments}

\clearpage
\hrule
\nocite{*}

\bibliographystyle{apsrev4-2}
\bibliography{apssamp}

\end{document}